\documentclass[fleqn,10pt]{wlscirep}
\usepackage[utf8]{inputenc}
\usepackage[T1]{fontenc}
\usepackage{colortbl}
\title{Export complexity, industrial complexity and regional economic growth in Brazil}

\author[1]{Ben-Hur Francisco Cardoso}
\author[1]{Eva Yamila da Silva Catela}
\author[1,2]{Guilherme Viegas}
\author[2,*]{Flávio L. Pinheiro}
\author[1]{Dominik Hartmann}

\affil[1]{Departamento de Economia e Relações Internacionais, Universidade Federal de Santa Catarina, Florianópolis, SC, Brazil}
\affil[2]{NOVA Information Management School (NOVA IMS), Universidade Nova de Lisboa, Portugal}
\affil[*]{fpinheiro@novaims.unl.pt}

\begin{abstract}
Research on productive structures has shown that economic complexity conditions economic growth. However, little is known about which type of complexity, e.g., export or industrial complexity, matters more for regional economic growth in a large emerging country like Brazil. Brazil exports natural resources and agricultural goods, but a large share of the employment derives from services, non-tradables, and within-country manufacturing trade. Here, we use a large dataset on Brazil's formal labor market, including approximately 100 million workers and 581 industries, to reveal the patterns of export complexity, industrial complexity, and economic growth of 558 micro-regions between 2003 and 2019. Our results show that export complexity is more evenly spread than industrial complexity. Only a few -- mainly developed urban places -- have comparative advantages in sophisticated services. Regressions show that a region’s industrial complexity is a significant predictor for 3-year growth prospects, but export complexity is not. Moreover, economic complexity in neighboring regions is significantly associated with economic growth. The results show export complexity does not appropriately depict Brazil's knowledge base and growth opportunities. Instead, promoting the sophistication of the heterogeneous regional industrial structures and development spillovers is a key to growth.
\end{abstract}
\begin{document}

\flushbottom
\maketitle%
\thispagestyle{empty}

\section*{Introduction}
Geographical differences in productivity and growth prospects can be largely explained by differences in the set of activities carried out at each location. In other words, what a region produces matters to achieve certain levels of income and well-being. Based on this idea, several measures of Economic Complexity -- such as the ECI \cite{hidalgo2009building} or the Fitness \cite{morrison2017economic} index -- have been proposed to characterize productive structures. These measures -- making use of dimensionality reduction techniques that capture the variety and ubiquity of an economy’s productive output -- are important determinants of economic growth \cite{hausmann2010country} and socioeconomic development \cite{hartmann2017linking}. The key underlying rationale is that it the expected economic return and level of value-added knowledge depends upon whether an economic agent (a country, region, firm, or worker) specializes in economic activities that many or few other agents can do.

Economic complexity literature has traditionally used international trade data at the country level to estimate country and product-level complexity indicators \cite{albeaik2017improving,morrison2017economic,hidalgo2009building}. However, it seems that export-based complexity measures are inadequate in comprehensively capturing knowledge and productive capabilities in all economic activities, such as non-tradeable services \cite{stojkoski2016impact}, and present significant biases when studying regional dynamics in continental-sized countries such as Brazil or the USA. Indeed, the data entry of exports might not necessarily be the actual production place of a good but instead the registry location of exports \cite{freitas2019industrias}. In recent years, researchers have turned their focus to patents \cite{balland2017geography,balland2018smart,petralia2017climbing,pinheiro2022dark}, academic output~\cite{guevara2016research,pugliese2019unfolding,stojkoski2023multidimensional}, and industry \cite{gao2021spillovers, fritz2021economic, chavez2017economic, pinheiro2022dark, freitas2023dataviva,koch2021yet} data to estimate the complexity of economies. Like exports, these datasets capture the intensity of an activity (e.g., exports, patents, academic production, industries, etc.) at different regional units (e.g., cities, municipalities, regions, or countries) to estimate the complexities of activities and regions. Hence, given the multiple ways one can estimate regional complexity, it is natural to ask which is the most appropriate measure of economic complexity. 

Research on regional complexity in developed economies (e.g., Europe and North America) tends to focus on patents to study technology complexity~\cite{balland2017geography,balland2018smart,pinheiro2022dark}, which seems natural as many of these countries can be considered at the technology frontier~\cite{klinger2006diversification} and compete for adoption or leadership in new technological domains. However, the same cannot be said to the same extent for regions in emerging economies, where patent production is regionally sparse. Hence, a substantial share of development research on Brazil and Latin America still focuses on trade data~\cite{operti2018dynamics,britto2019great,ferraz2018economic,bandeira2021economic,hartmann2021did}. However, it makes a difference if the policy focus is on protecting and promoting export sectors (e.g., via tariffs or better international port infrastructure) or whether the focus is on issues such as knowledge-based business sectors or within-country manufacturing trade (e.g., via better ICT infrastructure, roads, service spaces).

While recent works have investigated economic complexity based on industrial and occupational data \cite{fritz2021economic, pinheiro2022dark, zhu2020export, chavez2017economic, freitas2019industrias,de2022complexidade,freitas2023dataviva,queiroz2023economic,romero2021economic}, little emphasis has been put on exploring what is an adequate measure of economic complexity to depict regional development dynamics in developing and emerging economies, and how different measures adequately explain regional economic growth.

\begin{figure}[!t]
    \centering
    \includegraphics[width=0.95\textwidth]{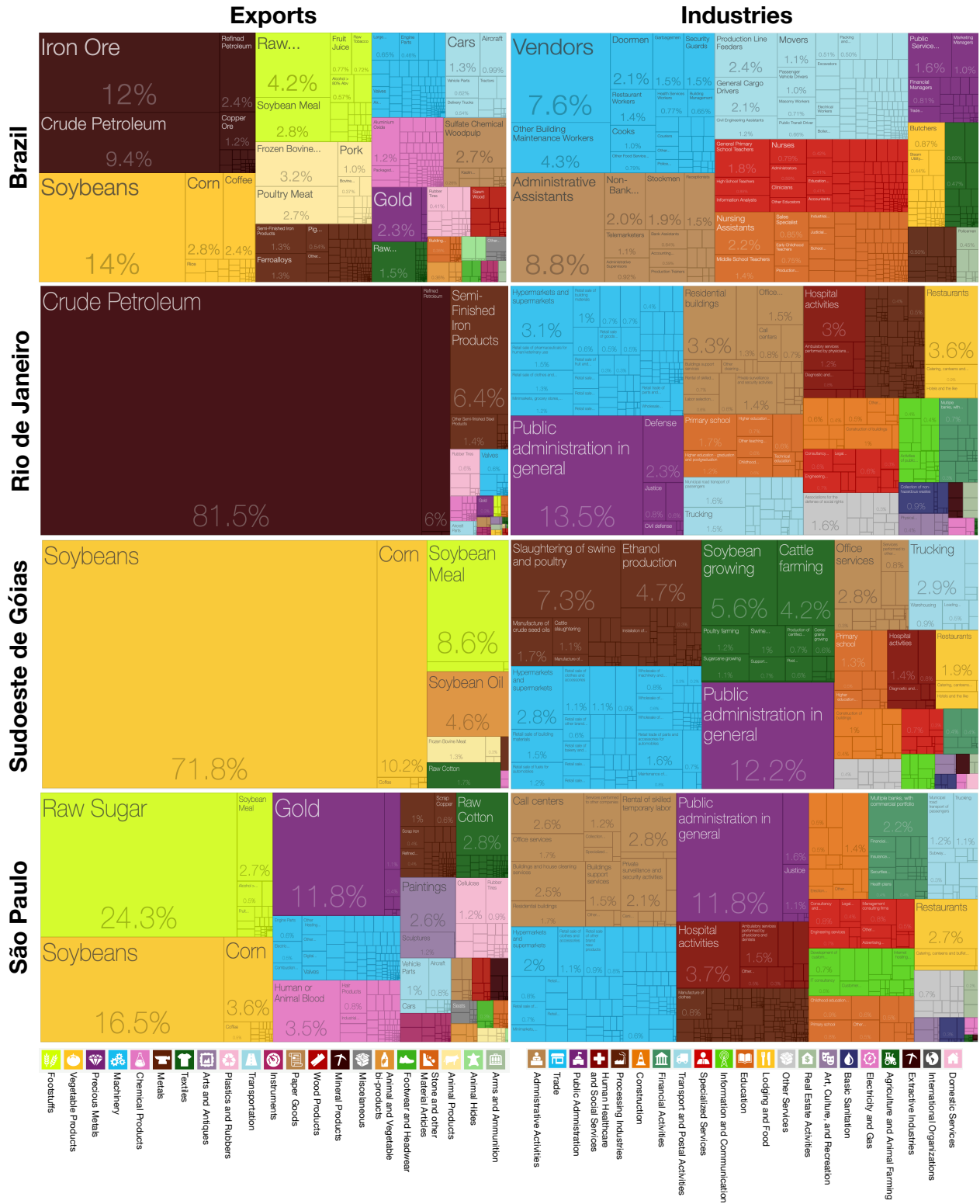}
    \caption{Export values in 2022 versus employment by industry data in 2021 for Brazil and the micro-regions Rio de Janeiro, the soybeans hub Sudoeste de Goias, and Sao Paulo. Brazil data at higher aggregation level ‘sections’ to better clarify product and industry types. Source, Dataviva.info \cite{dviva,hidalgo2014big}}
    \label{Figure1}
\end{figure}

Here, we estimate two measures of regional economic complexity across 558 micro-regions of Brazil using industry (IndECI) and export (ECI) data between 2003 and 2019. We then compare these Economic Complexity measures, showing that while the former shows a temporal tendency to increase agglomeration (i.e., build-up of spatial correlations), the latter does not. Still, they capture several stylized facts from economic complexity literature. Finally, we show that industrial complexity is more relevant than export complexity for regional economic growth. Nonetheless, regions also positively benefit from their neighboring regions' industrial and export complexity levels, arguably due to positive economic spillovers of demand, infrastructure, and knowledge. 

\section*{Data}
We use international trade data from the \textit{Growth Lab of Harvard University} \cite{dataverse}. The dataset covers trade between 242 countries between 2003 and 2019 in 1260 traded products classified according to the Harmonized System (HS) at the 4-digit level in the 1992 version. From COMEX, we sourced the FOB export value of each product by Brazilian municipality and month \cite{comex2015dados}. We aggregated the data to micro-regional \cite{IBGEa} and annual levels. From the \textit{Relação Anual de Informações Sociais} (RAIS), we make use of data that informs on the number of hours worked annually in each industry and municipality in Brazil between 2003 and 2019 within the formal labor market. The 581 industries are encoded using 5-digit version 2.0 of the \textit{Classificação Nacional de Atividades Econômica} (CNAE2.0). The data is aggregated at the micro-region level \cite{IBGEa}.

We did not explore the effects of patents and academic complexity for qualitative reasons. Patent applications are comparatively scarce~\cite{wipo} and thus seem inadequate for most regions and reduce our data considerably, apart from known problems of patents often being a measure of knowledge appropriation skills (and with a large share of foreign company applicants). Academic research output should be explored in more detail in the future. However, recent work on multidimensional economic complexity \cite{stojkoski2023multidimensional} points to a lower predictive power of research complexity than other ECI data sources. Moreover, it is known that research output in Brazil is mainly driven by public federal universities whose locations follow, to a significant extent, the population distribution as well as political reasons for broader access to higher education, and thus may also not depict the de-facto differences in productive capabilities well.

\begin{table}[!t]
\centering
\caption{Activities (products and industries) in the Bottom-5 and Top-5 of complexity estimated from exports (top) and industry (bottom), and the region with the highest RCA on such activity.}
\label{Table1}
\begin{tabular}{lll}
\hline
\rowcolor[HTML]{C0C0C0} 
\multicolumn{3}{l}{\cellcolor[HTML]{C0C0C0}\textbf{Product Complexity (Exports Data)}}                          \\ \hline
\rowcolor[HTML]{EFEFEF} 
product                                                      & PCI    & micro-region with highest RCA           \\ \hline
1801: Cocoa (whole or broken), raw or roasted                & -2.895 & Tomé-Açu, PA (RCA = 34.1)               \\
2609: Tin ores and concentrates                              & -2.874 & Ariquemes, RO (RCA = 1523.0)            \\
2602: Manganese ores \& concentrates inc mangnfrs iron ores  & -2.694 & Chorozinho, CE (RCA = 2680.2)           \\
2709: Crude oil from petroleum and bituminous minerals       & -2.636 & Baía da Ilha Grande, RJ (RCA = 18.2)    \\
5201: Cotton, not carded or combed                           & -2.606 & Tesouro, MT (RCA = 1270.3)              \\ \hline
...                                                          & ...    & ...                                     \\
9010: Apparatus and equipment for photographic…              & 2.094  & Barra do Piraí, RJ (RCA = 0.002)        \\
8461: Machine-tools for planing, shaping, slotting, …        & 2.104  & Itajubá, MG (RCA = 3.86)                \\
3707: Chemical preparations for photographic uses …          & 2.217  & Mogi das Cruzes, SP (RCA = 1.11)        \\
8457: Machining centres, unit construction machines …        & 2.411  & Sorocaba, SP (RCA = 3.50)               \\
3705: Photographic plates and film, exposed and …            & 2.468  & Jundiaí, SP (RCA = 2.99)                \\ \hline
\rowcolor[HTML]{C0C0C0} 
\multicolumn{3}{l}{\cellcolor[HTML]{C0C0C0}{\color[HTML]{000000} \textbf{Industry Complexity (Industry Data)}}} \\ \hline
\rowcolor[HTML]{EFEFEF} 
industry                                                     & ICI    & micro-region with highest RCA           \\ \hline
7227: Tin Ore Extraction                                     & -2.310 & Rio Preto da Eva, AM (RCA = 4408.5)     \\
84116: General Public Administration                         & -2.105 & Traipu, AL (RCA = 5.96)                 \\
7243: Precious Metal Ore Extraction                          & -2.056 & S. Miguel do Araguaia, GO (RCA = 228.6) \\
3213: Aquaculture in salty and brackish water                & -1.958 & Litoral Sul, RN (RCA = 403.7)           \\
47121: Retail trade of general merchandise ...               & -1.933 & Oiapoque, AP (RCA = 8.82)               \\
...                                                          & ...    & ...                                     \\
64328: Investment Banks                                      & 2.625  & São Paulo, SP (RCA = 7.57)              \\
64506: Capitalization Companies                              & 2.899  & Rio de Janeiro, RJ (RCA = 8.80)         \\
30504: Manufacturing of Military Combat Vehicles             & 3.521  & Campinas, SP (RCA = 49.54)              \\
66118: Administration of Exchanges and …                     & 3.671  & Osasco, SP (RCA = 8.92)                 \\
64409: Commercial Leasing                                    & 3.671  & Osasco, SP (RCA = 52.5)                 \\ \hline
\end{tabular}
\end{table}

Finally, we measure the micro-regional real GDP per capita (GDPpc) and micro-regional number of inhabitants (population) using data from \textit{Instituto Brasileiro de Geografia e Estatística} (IBGE) at the municipality level \cite{IBGEc} that was aggregated at the micro-regional level \cite{IBGEa}. We deflate the GDP to a 2010 base using the Brazilian Consumer Price Index \cite{IBGEb}. Unfortunately, there is no data on other attractive micro-regional controls, such as unemployment and the share of the population living in an urban area. Subsequently, we present our export- and industrial-based complexity measures in line with past works \cite{hidalgo2009building,pinheiro2022time}. It is noteworthy, though, is that the micro-regional level provides an analysis level due to conceptual reasons (designed by the statistical institute to capture connected labor markets).

\section*{Results and Discussion}
Brazil has an export structure strongly dependent on natural resources, such as soybeans, iron ore, or crude petroleum, see Figure \ref{Figure1}. However, when looking at employment data, many people in Brazil work in different types of processing industries or service activities. For example, the almost exclusive emphasis on Rio de Janeiro’s exports of crude petroleum does, of course, not properly depict the actual labor markets with many people working in different types of service activities, including public administration, commerce, and tourism, but also research and higher education activities, banking, etc (Figure \ref{Figure1}). São Paulo’s exports depict strengths in the export of cars, manufacturing, and chemical industries, but of course, a substantial share of these activities go to national clients. São Paulo is a hub for knowledge-based services (Figure \ref{Figure1}). This raises the question of whether export complexity or industrial complexity captures productive capabilities and growth prospects in Brazil more adequately. 

We start our analysis by comparing the industrial (ICI and IndECI) and export (PCI and ECI) complexity indicators and their correlation with economic growth. Table \ref{Table1} shows several services featuring among the highest- and lowest-ranked products and industries regarding the respective complexity indicators. Agricultural and mining activities -- such as simple commerce stores, herbs, and cocoa production -- and simple services (\textit{e.g.}, retail) are ubiquitous across Brazilian micro-regions and, thus, score low in complexity. At the same time, several types of specialized financial services -- such as investment banks or non-life insurance -- and different manufacturing activities -- such as vehicle parts, motors, or electronics -- are less ubiquitous and thus achieve a high complexity score. The industrial and export complexity indices rank manufacturing activities as high complexity and most agricultural activities as low complexity. 

\begin{figure}[!t]
    \centering
    \includegraphics[width=\textwidth]{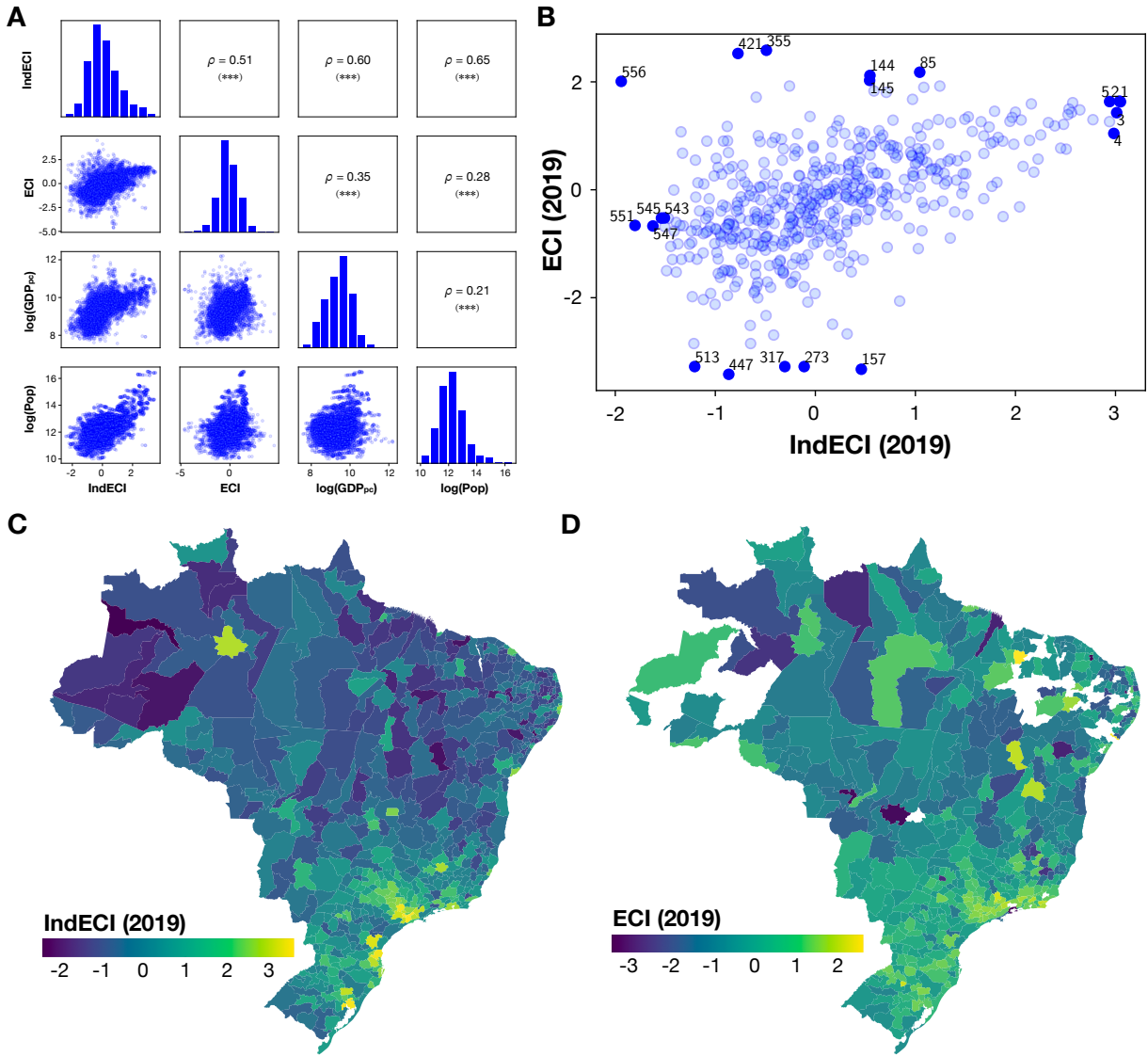}
    \caption{Panel A, Correlations between industrial complexity, export complexity, GDPpc, and population size of micro-regions. Panel B, Relationship between industry-based and export-based economic complexity index for 2019; the numbers correspond to the position in the IndECI ranking (Table~\ref{Table2}). In darker blue, the top and bottom five micro-regions of both indicators. The numbers correspond to the position in the IndECI ranking. Panel C, the spatial distribution of industry-based economic complexity index for 2019. Panel D, Spatial distribution of export-based economic complexity index for 2019. }
    \label{Figure2}
\end{figure}

A more substantial difference between the industrial complexity index and the export complexity index rankings emerges when we look at their association with other variables, such as GDP and population. Only an intermediate correlation of $\rho=0.51$ can be found between the industrial complexity and the export complexity index of the micro-regions (see Figure \ref{Figure2}A for a general correlation matrix and Figure \ref{Figure2}B for a more detailed scatter), and Table \ref{Table2} shows that there is no match between the top and bottom five micro-regions in both rankings. It is noteworthy that more well-known places such as São Paulo, Campinas, or Porto Alegre appear among the top industrial complexity regions but are absent in the exports-based rank. 

There is a higher correlation between industrial complexity and GDP per capita ($\rho=0.60$) than between export complexity and GDP per capita ($\rho=0.51$). A significantly stronger association can be observed between industrial complexity and population size ($\rho=0.65$) than between export complexity and population size ($\rho=0.21$). 

Both regional economic complexity indices tend to be higher in Brazil's Southern and Southeastern regions than in the Northern and North-Eastern regions, with the notable exception of several relatively complex larger cities in the latter regions. However, the spatial maps also illustrate that export complexity values are more dispersed than industrial complexity, which is more centered around metropolitan areas, such as São Paulo, Porto Alegre, Manaus, or Joinville, and more skewed towards the population denser coastal areas than the export complexity index (see Figures \ref{Figure2}C and \ref{Figure2}D). 

This points to a problem of the export complexity index (ECI) as an indicator for knowledge intensity in a country with large levels of spatial heterogeneity and inequality like Brazil, in which very few regions (such as Santa Rita do Sapucaí or Não-Me-Toque) de facto produce and focus on high export complexity goods. But even export cluster regions such as Santa Rita do Sapucaí focus on “Glass mirrors” and have an export complexity index of $2.18$, or Não-Me-Toque, focuses on the production of “Agricultural machinery for soil preparation or cultivation” and an ECI index of $2.12$. For many regions in Brazil, it is enough to export intermediate to low-level export goods like pork and chicken meat or textile products to be considered relatively complex compared to regions that export none or even simpler products. 

\begin{table}[]
\centering
\caption{Top and bottom regions based on their industry-based and export-based economic complexity index in 2019. The numbers in parentheses correspond to the position in the IndECI ranking. The list only contains micro-regions that have exports (and, therefore, an ECI value)}
\label{Table2}
\begin{tabular}{lllll}
\multicolumn{2}{c}{\textbf{IndECI ranking}}                           &  & \multicolumn{2}{c}{\textbf{ECI ranking}}                           \\ \cline{1-2} \cline{4-5} 
\cellcolor[HTML]{EFEFEF}IndECI & \cellcolor[HTML]{EFEFEF}Micro-region &  & \cellcolor[HTML]{EFEFEF}ECI & \cellcolor[HTML]{EFEFEF}Micro-region \\ \cline{1-2} \cline{4-5} 
3.0496                         & (1) Osasco                           &  & 2.5853                      & (355) Codó                           \\
3.0405                         & (2) Guarulhos                        &  & 2.5256                      & (421) Propriá                        \\
3.0120                         & (3) São Paulo                        &  & 2.1786                      & (85) Santa Rita do Sapucaí           \\
2.9832                         & (4) Porto Alegre                     &  & 2.1187                      & (144) Não-Me-Toque                   \\
2.9411                         & (5) Campinas                         &  & 2.0276                      & (145) Bacia de São João              \\
...                             & ...                                   &  & ...                          & ...                                   \\
-1.5087                        & (543) Caracaraí                      &  & -3.2764                     & (513) Tesouro                        \\
-1.5386                        & (545) Boca do Acre                   &  & -3.2764                     & (273) Alto Paraguai                  \\
-1.6240                        & (547) Bico do Papagaio               &  & -3.2764                     & (317) Esperança                      \\
-1.8027                        & (551) Chapadas das Mangabeiras       &  & -3.3258                     & (157) Baía da Ilha Grande            \\
-1.9403                        & (556) Cotegipe                       &  & -3.4197                     & (447) Chorozinho                     \\ \cline{1-2} \cline{4-5} 
\end{tabular}
\end{table}

While this might make sense from a “first steps of industrialization” perspective, it is also slightly at odds with a modern understanding of a knowledge-based society in which high-skilled workers work in knowledge-based services activities in urban centers with a high density of higher education facilities, access to diverse ideas, and a differentiated demand structure~\cite{jacobs2016death,florida2002rise,glaeser1992growth,giovanini2023productive}. 

\begin{figure}[!t]
    \centering
    \includegraphics[width=\textwidth]{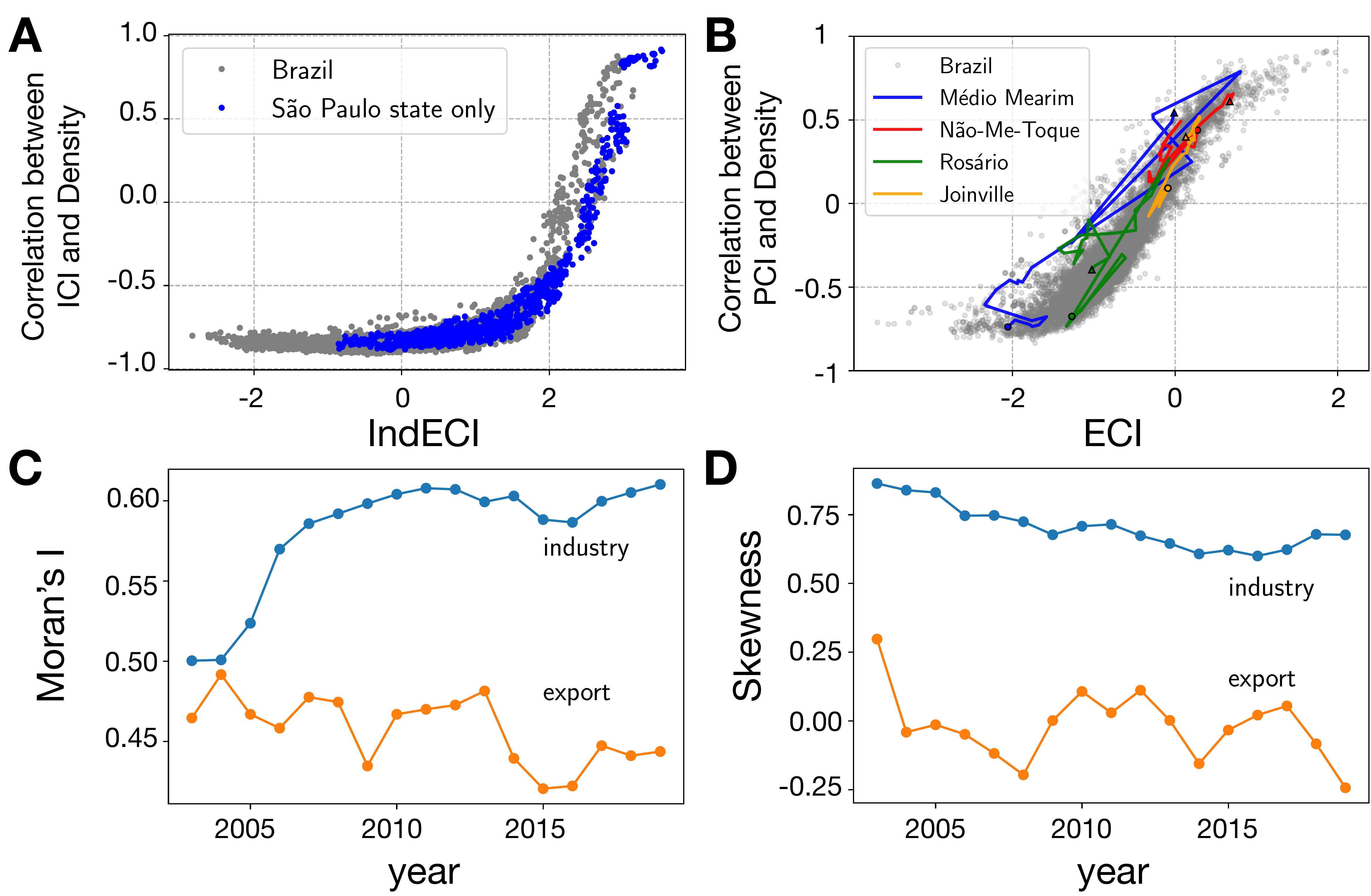}
    \caption{The S-curve of economic development for both industry-based (Panel A) and export-based (Panel B) economic complexity indexes depicting the closeness of each micro-region to complex activities given their complexity level. Panel C shows the time evolution of spatial autocorrelation (Moran’s I), and Panel D shows the spatial concentration (skewness) of both complexity metrics. }
    \label{Figure3}
\end{figure}

To further understand the distribution of industrial and export complexity, we plot in Figure \ref{Figure3}A and \ref{Figure3}B the regional complexity indices on the horizontal axis versus the closeness of the regions to new complex activities on the Y-axis. Previous work on this usually S-shaped association/curve of productive sophistication \cite{pinheiro2018shooting,pinheiro2022dark,pinheiro2022time,hartmann2020international,hartmann2021did} showed that countries and regions tend to first diversify in the initial stages of development first into related, relatively simple activities, and only at intermediate to high levels of economic development move closer to complex activities. This implies a certain gravitation of regions with simple productive structures towards diversification into simple activities and of complex regions towards complex activities \cite{hidalgo2007product,hausmann2010country}. 

We show a more linear association between the export complexity index (ECI) and the closeness to export products across micro-regions of Brazil than between industry complexity (IndECI) and closeness to complex activities. There is a very steep slope beyond a relatively high industrial complexity index. Only a few Brazilian micro-regions (mainly some urban centers) are close to the most complex industries. Moreover, it is noteworthy that industrial complexity evolution is more stable than export complexity evolution, and a significant variation in the growth rates of industrial versus export complexity between 2003 and 2019 can be observed. Regions like Joinville illustrate a certain correlation in the trend of both variables. However, in other cases (such as Médio Mearim) big jumps have been made in export complexity, but little progress has been made in industrial complexity. Initial levels of industrial complexity are different, more stable, and better predictors of future economic development trajectories.

Regarding the spatial distribution of complexity indices. Figure \ref{Figure3}C shows the Moran's I Index \cite{moran1950notes}, which is computed as

\begin{equation}
    \text{Morans' I} = \frac{||R||}{\sum_{r\in R}||N(r)||}\frac{\sum_{r \in R}\sum_{r' \in N(r)} (\text{ECI}_{r'}-\langle \text{ECI} \rangle)(\text{ECI}_r-\langle \text{ECI} \rangle)}{\sum_{r \in R} (\text{ECI}_r-\langle \text{ECI} \rangle) ^2} 
\end{equation}
\noindent where $R$ is the set of 558 micro-regions and N(r) is the set of neighbor micro-regions of micro-region $r$. The values of Morans' I range from -1 to +1 and quantify spatial autocorrelations. A negative spatial autocorrelation is associated with patterns like a chessboard (in which white squares surround all black squares and vice versa) and positive autocorrelation with regions neighboring other regions that are similar. A random arrangement would lead to a Moran's I value close to 0. 

Moreover, regarding the distribution of the complexity indexes, Figure 3D shows the right-skewness, which can be estimated as
\begin{equation}
    \text{skewness} = \frac{\langle(\text{ECI}_r-\langle \text{ECI}\rangle)^3\rangle}{\langle(\text{ECI}_r-\langle \text{ECI}\rangle)^2\rangle^\frac{3}{2}}
\end{equation}

Figure \ref{Figure3}D shows that, during the analysis period, industrial complexity (IndECI) exhibits consistently higher levels of right-side skewness than export complexity (ECI), which indicated that IndECI is shifted towards higher values on the right side of the distribution. Additionally, as shown in Figure \ref{Figure3}C, IndECI exhibits higher spatial autocorrelations, measured in terms of the Morans' I index, than export complexity (ECI). This means that neighboring regions more often have similar values in terms of industry complexity than export complexity. However, in both cases, spatial spillover effects are likely to affect the economic behavior of the micro-regions. Notably, the spatial autocorrelations of regional export complexity have slightly reduced from 0.46 in 2003 to 0.44 in 2019, and the spatial autocorrelation of industry complexity has significantly gone up from 0.50 in 2003 to 0.61 in 2019.

\subsection*{Regional Complexity and Economic Growth}
Next, we analyse the role of each economic complexity indicator in the economic growth prospects of each region. We quantify how much the economic complexity of a focal region and its neighboring regions impact the economic growth of the focal region. To that end, we define the average complexity of the neighbors of a focal region simply as:
\begin{equation}
    \langle \text{ECI} \rangle ^t_{N(r)} = \frac{1}{||N(r)||}\sum_{r'\in N(r)} \text{ECI}^t_{r'}
\end{equation}

We then follow with a regression analysis to examine the link between the economic growth of a focal region with its industry- and export-based economic complexity and the average complexity of their geographical neighbors. We also control the micro-regional real GDP per capita ($y_r^t$) and micro-regional population size ($N_r^t$). We define the continuous-time-equivalent 3-year growth rate as $g_r^t = (\text{log} y_r^{t+3} - \text{log} y_r^t)/3$. Results are robust considering 2- and 4-year regional growth, as shown in the SI. 

In the end, we focus our analysis on the following model that regresses economic growth ($g_r^t$) against a set of relevant independent variables (described above), such that:
\begin{equation}
    \begin{split}
        g_r^t = & \beta_0 + \beta_1\text{IndECI}_r^t+\beta_2 \langle\text{IndECI}\rangle_{N(r)}^t + \beta_3 \text{ECI}_r^t +  \beta_4 \langle\text{ECI}\rangle_{N(r)}^t + \\
        & \beta_5 \text{log}y_r^t + \beta_6 \text{log} N_r^t + \rho g_r^{t-1} +  \mu_r + \nu^t + \epsilon_r^t
    \end{split}
\end{equation}
\noindent where $\mu_r$ and $\nu^t$ are regional and year-fixed effects, and the residuals $\epsilon_r^t$ are assumed to be exogenous. Because we control these fixed effects, our model can capture the effect of different types of time-invariant characteristics of micro-regions and nationwide trends with year-fixed effects.

\begin{table}[!t]
\centering
\caption{Results of the regression model showing estimates of the average marginal effect of the local economic complexity index with neighbors’ terms.}
\label{Table3}
    \begin{tabular}{lcccccccc}
    & \multicolumn{8}{c}{GDP growth, $g^t$}   \\ \hline
     & (1)  & (2)   & (3)   & (4)   & (5)   & (6)   & (7)   & (8)   \\ \hline
    IndECI     && & \begin{tabular}[c]{@{}l@{}}0.0987$^{***}$\\ (0.0198)\end{tabular}  & & \begin{tabular}[c]{@{}l@{}}0.0427$^{***}$\\ (0.0147)\end{tabular}  &  & & \\
    $\langle \text{IndECI} \rangle_N$ && & & \begin{tabular}[c]{@{}l@{}}0.1526$^{***}$\\ (0.0173)\end{tabular}  & \begin{tabular}[c]{@{}l@{}}0.1056$^{***}$\\ (0.0147)\end{tabular}  & & & \\
    ECI  && & & & & \begin{tabular}[c]{@{}l@{}}0.0041\\ (0.0092)\end{tabular}     & & \begin{tabular}[c]{@{}l@{}}0.0044\\ (0.0071)\end{tabular}     \\
    $\langle \text{ECI} \rangle_N$    && & & & & & \begin{tabular}[c]{@{}l@{}}0.1330$^{***}$\\ (0.0164)\end{tabular}  & \begin{tabular}[c]{@{}l@{}}0.1199$^{***}$\\ (0.0157)\end{tabular}  \\
    log $y$    && \begin{tabular}[c]{@{}l@{}}-0.4789$^{***}$\\ (0.0501)\end{tabular} & \begin{tabular}[c]{@{}l@{}}-0.4182$^{***}$\\ (0.0369)\end{tabular} & \begin{tabular}[c]{@{}l@{}}-0.3771$^{***}$\\ (0.0322)\end{tabular} & \begin{tabular}[c]{@{}l@{}}-0.3254$^{***}$\\ (0.0266)\end{tabular} & \begin{tabular}[c]{@{}l@{}}-0.4406$^{***}$\\ (0.0434)\end{tabular} & \begin{tabular}[c]{@{}l@{}}-0.3593$^{***}$\\ (0.0337)\end{tabular} & \begin{tabular}[c]{@{}l@{}}-0.3357$^{***}$\\ (0.0313)\end{tabular} \\
    log $N$    && \begin{tabular}[c]{@{}l@{}}-0.2672$^{***}$\\ (0.0702)\end{tabular} & \begin{tabular}[c]{@{}l@{}}-0.2140$^{***}$\\ (0.5059)\end{tabular} & \begin{tabular}[c]{@{}l@{}}-0.2025$^{***}$\\ (0.0504)\end{tabular} & \begin{tabular}[c]{@{}l@{}}-0.1299$^{***}$\\ (0.0339)\end{tabular} & \begin{tabular}[c]{@{}l@{}}-0.2271$^{***}$\\ (0.0598)\end{tabular} & \begin{tabular}[c]{@{}l@{}}-0.2234$^{***}$\\ (0.0565)\end{tabular} & \begin{tabular}[c]{@{}l@{}}-0.2020$^{***}$\\ (0.0502)\end{tabular} \\
    $g^{t-1}$  & \begin{tabular}[c]{@{}l@{}}0.6393$^{***}$\\ (0.0320)\end{tabular} & \begin{tabular}[c]{@{}l@{}}0.3663$^{***}$\\ (0.0450)\end{tabular}  & \begin{tabular}[c]{@{}l@{}}0.4113$^{***}$\\ (0.0365)\end{tabular}  & \begin{tabular}[c]{@{}l@{}}0.4062$^{***}$\\ (0.0333)\end{tabular}  & \begin{tabular}[c]{@{}l@{}}0.4447$^{***}$\\ (0.0301)\end{tabular}  & \begin{tabular}[c]{@{}l@{}}0.3947$^{***}$\\ (0.0387)\end{tabular}  & \begin{tabular}[c]{@{}l@{}}0.4151$^{***}$\\ (0.0319)\end{tabular}  & \begin{tabular}[c]{@{}l@{}}0.4312$^{***}$\\ (0.0302)\end{tabular}  \\
    Constant   & \begin{tabular}[c]{@{}l@{}}0.0931$^{***}$\\ (0.0063)\end{tabular} & \begin{tabular}[c]{@{}l@{}}7.7413$^{***}$\\ (1.0896)\end{tabular}  & \begin{tabular}[c]{@{}l@{}}6.5252$^{***}$\\ (0.7819)\end{tabular}  & \begin{tabular}[c]{@{}l@{}}5.9851$^{***}$\\ (0.739)\end{tabular}   & \begin{tabular}[c]{@{}l@{}}4.6335$^{***}$\\ (0.5153)\end{tabular}  & \begin{tabular}[c]{@{}l@{}}6.9039\\ (0.9141)\end{tabular}     & \begin{tabular}[c]{@{}l@{}}6.4831$^{***}$\\ (0.8511)\end{tabular}  & \begin{tabular}[c]{@{}l@{}}5.9746$^{***}$\\ (0.7619)\end{tabular}  \\ \hline
    Fixed Effects \\
    Micro-region    & Yes  & Yes   & Yes   & Yes   & Yes   & Yes   & Yes   & Yes   \\
    Year  & Yes  & Yes   & Yes   & Yes   & Yes   & Yes   & Yes   & Yes   \\ \hline
    AB test    & \checkmark    & \checkmark     & \checkmark     & \checkmark     & \checkmark     & \checkmark     & \checkmark     & \checkmark     \\
    Sargan Test& \checkmark    & \checkmark     & \checkmark     & \checkmark     & \checkmark     & \checkmark     & \checkmark     & \checkmark     \\ \hline
    Observations   &5773&5773&5773&5773&5773&5773&5773&5773\\ \hline
    \end{tabular}
\end{table}

To detect the presence of autocorrelation, heteroscedasticity, multicollinearity, and endogeneity of the variables (Allano and Bond, Breusch-Pagan), we assessed the variance inflation factor (VIF) and performed a Durbon-Wu-Hausman tests. Based on these tests, it was verified that the dataset used presents the following characteristics that must be treated so that the estimates are consistent: a) the dependent variable is dynamic, in the sense that it depends on its past values; b) the independent variables are not strictly exogenous; c) there are fixed effects in observational units (micro-regions) and time periods, which must be controlled; d) multicollinearity between ECI and IndECI.

Hence, considering regional and temporal fixed effects, a dynamic Arellano–Bover/Blundell–Bond panel (Rodman, 2009) was used. This estimator is appropriate for the data set used, given that we have: a) a greater number of units than years; b) we consider a linear functional relation; c) there is an autoregressive coefficient; d) the independent variables are not strictly exogenous; e) there is a regional and temporal fixed effect. After estimation, both a Sargan and Sargan-Hansen test reject the null hypothesis of over-identification of the instruments and the validity of the instruments used. The Arellano-Bond test allows for rejecting the null hypothesis of serial correlation.

Table \ref{Table3} shows the results of the regression model. The first two columns show the model results with only the control variables. In the case of Column 1, the three-year lagged average growth rate is positive and significant, as expected. When we include GDP per capita and population size in Column 2, the result of the lagged dependent variable remains significant and with the same sign. Including these control variables is justified to absorb the scale of the region and the richness.

The complexity variable IndECI Is included in column 3. As expected, the coefficient is positive (and significant), which indicates that the region's growth rate is positively related to a higher industrial complexity indicator. On the other hand, in Column 4, it is noticed that there are neighborhood spillovers, given that a higher indicator of $\langle \text{IndECI}\rangle_N$ of the neighbors of each micro-region is also positively related to the average growth rate. In Column 5, the two variables IndECI and $\langle \text{IndECI}\rangle_N$ are combined, and the results remain significant. However, we have a decrease in the complexity coefficient of the region, which is lower than the complexity indicator of the neighbors. Columns 6, 7, and 8 present the results when considering the variables related to exports from the micro-region and close neighbors. Only the second of these variables was significant and with the expected sign.

Thus, an increase in the export complexity of neighboring regions is significantly associated with an increase in the economic growth prospects of the target region. Still, an increase in export complexity within the region is insufficient to raise GDP per capita significantly (at least within the analysed time period).

\begin{figure}[!t]
    \centering
    \includegraphics[width=\textwidth]{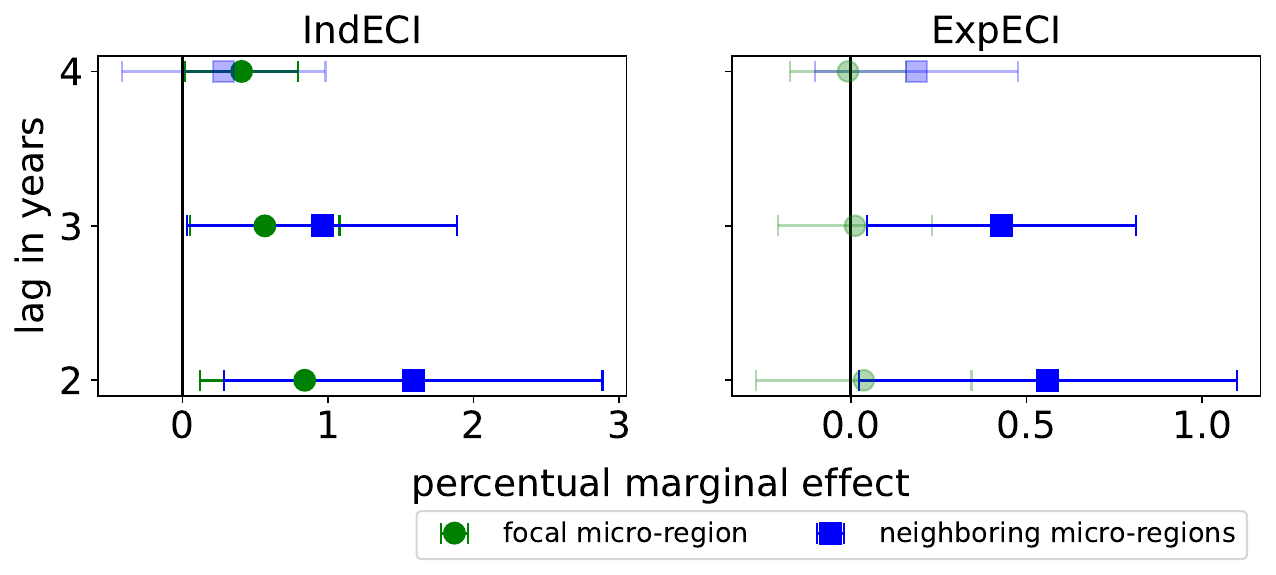}
    \caption{The marginal effect in different time lag growth.}
    \label{Figure4}
\end{figure}

To test the robustness of the thesis of this work, we considered different time intervals to measure the growth of micro-regions. As Figure \ref{Figure4} shows, we see that the results are quite general: The industrial complexity of a location and its neighborhood are important for growth, while only the complexity of the neighborhood's exports matters for a location's growth. The regression table for all these results can be found in the SI.

These results imply that an economic diversification strategy that only considering exports may not be sufficient to use local knowledge, smart diversification, and growth potential in a large developing/emerging economy such as Brazil. Instead of identifying with complete industrial data, for instance, significant opportunities for regions to move into knowledge-based services or manufacturing activities for the large Brazilian market, export data can lead to identifying relatively simple, natural resource-exploiting goods\cite{hartmann2023complexidade}. This could further perpetuate the traditional problems of external dependency and focus on comparative advantages in natural resources, agricultural, and mining activities instead of structural transformation into more knowledge-based and value-added activities~\cite{furtado2020formaccao,santos1970structure,hartmann2020international}.

\section*{Conclusions}
Initial approaches from economic complexity research have tended to focus on export structures as indicators of productive structures~\cite{hidalgo2009building,hidalgo2021economic}, as it best fits available data and it is in line with literature highlighting the problem of core-periphery structures in trade and the problem of external dependence of countries~\cite{prebisch1959commercial,santos1970structure,gala2018economic,hartmann2020international}. The application of complexity indicators to regions of developed economies have tended to use patent data as a proxy for the technology frontier\cite{pinheiro2022time,balland2018smart}. 

In this article, we show that industrial complexity is more adequate at capturing the knowledge base and predicting economic growth at the micro-regional level in a large emerging economy, such as Brazil, than export data. Export data does not capture significant strengths of regions in services as well as manufacturing activities destined for the large domestic market. Indeed, a large share of the workforce does not work in export-sectors and thus export data strongly underestimates the actual knowledge and skills of the workforce.

Our results show a stronger spatial concentration and skewness of industrial complexity values than export complexity values in Brazil. Only a few urban hubs have managed to achieve comparative advantages in the most complex services and specialized manufacturing activities, while low to intermediately complex agro-industrial exports are already sufficient to achieve relatively high economic complexity values in the within-country export complexity comparisons. Industrial complexity is a strong predictor of regional economic growth, however, the export complexity of a region is not, at least in the case of Brazil. 

These results invite a rethinking of complexity as a policy tool in the context of regional development policies. Indeed, complexity-based indicators are multidimensional and highly dependent on the choice of underlying proxy data. In effect, they capture different dimensions of productive systems and knowledge bases, which arguably should be evaluated as separate, complementary, and/or interrelated challenges in future work. Simply put, increasing the export complexity of a region does not directly entail improving the region’s industry complexity, or vice versa. In the case of countries, such as Brazil, a mere emphasis on exports is not sufficient to lift all regional boats, and endogenous smart diversification opportunities are not identified.  This does not mean that an increase in more knowledge- and technology-intensive exports should not be part of smart diversification strategies of Brazil. Yet it is not sufficient to ensure economic growth across the heterogeneous regions of large emerging economies, such as Brazil. 

This study shed some light on the complicated matter of finding the right economic complexity dimension to promote economic growth across the regions of a large emerging economy. Future research arguably needs to benchmark results from countries at different stages of economic development and with different domestic market sizes.

\section*{Methods}
\subsection*{Exports and Industry Complexity}
Let $X_{r,i}^{t}$ measure the revealed intensity of an activity $i$ in region $r$ on year $t$. We defined the Revealed Comparative Advantage (RCA) of region $r$ in activity $i$ as 
\begin{equation}
    \text{RCA}_{r,i}^t = \frac{X_{r,i}^{t}}{\sum_j X_{r,j}^{t}} \frac{1}{Z_i^t}
\end{equation}
the upper term measures the relative intensity of activity $i$ in the region $r$, and $Z_i^t$ provides a baseline expectation of the relative intensity of such activity from a typical region. This is equivalent to the Location Quotient and the Balassa Index\cite{hoen2006measurement}.

We estimate the RCA using both industry and export intensity. In that sense, for industries, we measured intensity through the number of hours worked in the micro-region, and the baseline corresponds to the share of the number of hours spent in an industry $i$ across the entire country, that is, 
\begin{equation}
    Z_i^t = \frac{\sum_{r'} X_{r',i}^{t}}{\sum_{r',i'} X_{r',i'}^{t}}
\end{equation}
For exports, we measure intensity through the total reported exports in USD of a region in a given product, and, following previous works \cite{simoes2011economic,hidalgo2021economic}, we consider the product's share in global international trade between countries as a baseline ($Z_i^t$). 

An RCA greater/lower than one implies that a region reveals an intensity on a given activity more/less than what we expect from a typical region. Next, we can represent the regional division of activities through a specialization matrix $M^t$, with entries $M_{r,i}^t$  equal to $1$ if the micro-region $r\in R$  reveals a significant intensity of activity $i$ in the year $t$ ($\text{RCA}_{r,i}^t > 1$) and $0$ otherwise.

Following standard methods from economic complexity\cite{hidalgo2021economic}, the complexity of a micro-region is the average complexity of the activities in which it has specialized in. As such, we can use the Product Complexity Index (PCI) estimated from international trade data\cite{hidalgo2021economic} to obtain the complexities of each exported product and use the average to estimate each region’s export Economic Complexity Index (ECI):
\begin{equation}
    \text{ECI}_r^t = \frac{1}{M_{r,\ast}^t}\sum_i M_{r,i}^t\text{PCI}_i^t
\end{equation}

However, we need to estimate both quantities from the existing industry data. As such, consider, 
\begin{equation}
    K_r^t = \frac{1}{M_{r,\ast}^t}\sum_i M_{r,i}^tQ_i^t
\end{equation}
as the complexity of a micro-region ($K_r^t$) given the complexities of industries ($Q_i^t$). Conversely, the complexity of activity ($Q_i^t$) is the average of the complexity of the micro-regions ($K_r^t$) specialized in it:
\begin{equation}
    Q_i^t = \frac{1}{M_{\ast,i}^t}\sum_r M_{r,i}^tK_r^t
\end{equation}
which, after some manipulations, leads to:
\begin{subequations}
    \begin{align}
             K_r^t = & \sum_{r'} \hat{M}_{r,r'}^tK_{r'}^t \\ 
             \hat{M}_{r,r'} = & \sum_i \frac{M_{r,i}^tM_{r',i}^t}{M_{r,\ast}^tM_{\ast,i}^t}
    \end{align}
\end{subequations}
and
\begin{subequations}
    \begin{align}
             Q_i^t = & \sum_j \Tilde{M}_{i,j}^tQ_{j}^t \\ 
             \Tilde{M}_{i,j} = & \sum_i \frac{M_{r,i}^tM_{r,j}^t}{M_{r,\ast}^tM_{\ast,i}^t}
    \end{align}
\end{subequations}
To solve these recursive equations, it is enough to identify that the complexity vector of the regions $K^t$ is an eigenvector of $\hat{M}_t$ and the complexity vector of the activities $Q^t$ is an eigenvector of $\Tilde{M}_t$, both associated with the second largest eigenvalue since it captures the largest amount of variation in the system. As $K_r^t$  and $Q_i^t$ are relative metrics \cite{fritz2021economic}, the industry-based complexity of micro-regions (IndECI) and the Industry Complexity Index (ICI) are defined after applying the following standardization:
\begin{subequations}
    \begin{align}
             \text{IndECI}_r^t = & \frac{K_r^t-\langle K^t\rangle}{\text{std}(K^t)} \\
             \text{ICI}_i^t = & \frac{Q_i^t-\langle Q^t\rangle}{\text{std}(Q^t)}
    \end{align}
\end{subequations}

\subsection*{Related Density and Closeness to Complex Activities}
We measure the density $\omega$ of an activity with revealed comparative advantages around activity $i$ in the portfolio of activities of region $r$ as
\begin{equation}
    \omega_{ri} = \frac{\sum_{i'}M_{ri'}\phi_{ii'}}{\sum_{i'}\phi_{ii'}}
\end{equation}
where $\phi_{ii'} = \text{min}(P(\text{RCA}_{r,i}\geq 1 |\text{RCA}_{r,i'}\geq 1), P(\text{RCA}_{r,i'}\geq 1 |\text{RCA}_{r,i}\geq 1)$ measures the proximity between two activities and is measured as the minimum conditional probability that a region has revealed comparative advantage in both activities. The proximities form the backbone of the Product Space\cite{hidalgo2007product} and the Industry Space~\cite{gao2021spillovers}. Relatedness is a well-known predictor of the ability of regions to develop new activities\cite{hartmann2021did}, implying that regions are more likely to enter activities that are related to their current capabilities.

In order to analyse the distance of each region’s economic structure to simple/complex new activities, we measure the Pearson correlation between the measured density ($\omega$) of activities without revealing comparative advantage in regions and the complexity of such activities (PCI or ICI) of these potential new activities. In that sense, a negative correlation indicates that the diversification opportunities of regions are mainly simple activities, a positive correlation indicates that the closest diversification opportunities are mainly complex activities, a null correlation indicates that the region is equally close to simple and complex activities\cite{pinheiro2018shooting,hartmann2019identifying}. 

In this measure of “closeness to complex activities,” it is important to mention that we consider only activities for which a region has not yet exhibited a revealed comparative advantage. Such choice has two main reasons: first, the focus of these measures is on the future development opportunities at different levels of economic development; secondly, measuring distance/closeness to existing activities ($M_{ri}=1$) is redundant as such activities already make up a region’s portfolio and thus contribute already to the regional complexity.

\section*{Acknowledgements}

BFC would like to express his gratitude for the financial support of the Coordenação de Aperfeiçoamento de Pessoal de Nível Superior (CAPES) – Finance Code 001, and DH for the support of CNPq (406943/ 2021-4 and 315441/2021-6). F.L.P. acknowledges the financial support provided by FCT Portugal under the project UIDB/04152/2020 -- Centro de Investigação em Gestão de Informação (MagIC). We are grateful for valuable comments of Diogo Ferraz, Marcelo Arend, Pablo Bittencourt, NECODE members, and participants at the Anpec Sul conference and research seminars at UFSC and FIESC in 2023. 

\section*{Author contributions statement}
B.F.C., E.Y.S.C, F.L.P., and D.H. conceived the experiment(s), B.F.C. and G.V. conducted the experiment(s), B.F.C., E.Y.S.C, and D.H. analysed the results.  
All authors participated in the writing and revision of the manuscript.

\section*{Additional information}
The authors declare no competing interests.

\end{document}